\documentclass[a4paper, headings=standardclasses, headings=big, captions=tableheading]{scrartcl}

\pdfoutput=1

\usepackage[utf8]{inputenc}
\usepackage[english]{babel}
\usepackage{csquotes}
\usepackage{gensymb}
\usepackage{pifont}

\usepackage{hyperref}
\addto\extrasenglish{%

}

\usepackage[dvipsnames]{xcolor}
\usepackage{supertabular}
\usepackage{graphicx}
\usepackage{caption}
\usepackage{booktabs}
\usepackage{rotating}

\usepackage[misc]{ifsym}

\usepackage[norule,bottom,ragged,hang,multiple]{footmisc}

\newcommand{\linebreakcell}[2][c]{%
	\begin{tabular}[#1]{@{}c@{}}#2\end{tabular}}

\newcommand{\newterm}[1]{\textbf{#1}}
\newcommand{\tocite}[1]{{\color{orange}[?]}}

\newcommand{\greencheck}{{\color{ForestGreen}\ding{51}}}
\newcommand{\redcross}{{\color{BrickRed}\ding{55}}}
\newcommand{\grayquestionmark}{{\color{gray}\textbf{?}}}

\usepackage{fancyhdr}
\newcommand{\setfooter}{\fancyfoot[C]{
	\thepage
}}

\fancypagestyle{plain}{
	
	\fancyhead[L]{}
	\fancyhead[R]{}
	\setfooter{}
}

\setfooter{}
\usepackage[backend=biber,sorting=ynt,urldate=long,block=ragged]{biblatex}
\title{Using CEF Digital Service Infrastructures in the Smart4Health Project for the Exchange of Electronic Health Records}
\author{
	\Large Tamara Slosarek$^1$
	\and
	\Large Attila Wohlbrandt$^1$
	\and
	\Large Erwin Böttinger$^{1,2}$
}
\date{
	{\small $^1$Digital Health Center, Hasso Plattner Institute, University of Potsdam,\\
	Prof.-Dr.-Helmert-Str. 2-3, 14482 Potsdam, Germany\\\vspace{0.25em}
		\Letter~\texttt{\{firstname.lastname\}@hpi.de}\\\vspace{0.75em}
		$^2$Hasso Plattner Institute for Digital Health at Mount Sinai,\\
		Ichan School of Medicine at Moint Sinai,\\
		1 Gustave L. Levy Pl, New York, NY 10029, USA\\}
	\vspace{\baselineskip}\today
}

\begin{document}

\maketitle

\begin{abstract}

The Smart4Health (S4H) software application will empower EU citizens to manage, analyze, and exchange their aggregated electronic health data. In order to provide such a service, basic features are needed to ensure usability, reliability, and trust. The CEF building blocks implement such functionalities while complying with EU regulations.
This report examines the current status and applicability of the CEF building blocks for the envisioned S4H software application.

The major findings of the report are that (1) most CEF building blocks are currently not ready to be applied without further implementation efforts and (2) the S4H-specific use cases and user needs, to which the single CEF building blocks correspond, must be clarified. Open questions raised in this report need to be answered before a clear analysis can be made. Moreover, the functionalities of CEF building blocks aim at the matured product, while for the first version of the S4H software application basic implementations suffice. Still, concepts need to be elaborated of how the sample implementation of CEF building blocks or suitable alternative applications can be included into the system at a later point.

\end{abstract}

\section{Introduction}

The Smart4Health (S4H) project~\autocite{smart4health} aims to create a citizen-centered platform for the management and exchange of health records across Europe.

For the development of a usable service dealing with confidential information in form of medical data, specific functionalities such as identification and authentication, translation, and a secure data transfer are needed, especially in a pan-European setup.
Therefore, the Connecting Europe Facility (CEF) programme~\autocite{cefHomepage} funds different projects that implement Digital Service Infrastructures (DSIs). More general DSIs, called CEF building blocks, such as eID and eTranslation, can be reused by other projects and larger DSIs, such as eHealth. CEF building blocks that implement trust services comply to European law as laid down in the eIDAS Regulation~\autocite{eIDAS}.

This report examines, which DSIs are adequate for the S4H software application and to what extend. The report describes relevant aspects of the S4H system architecture in \autoref{sec:s4h}, and CEF DSIs in the S4H context in \autoref{sec:cef_dsis}. \autoref{sec:conclusion} summarizes recommendations on the incorporation, based on the preceding sections.
\section{The S4H Architecture}
\label{sec:s4h}

\begin{figure*}[ht]
	\includegraphics[width=\textwidth]{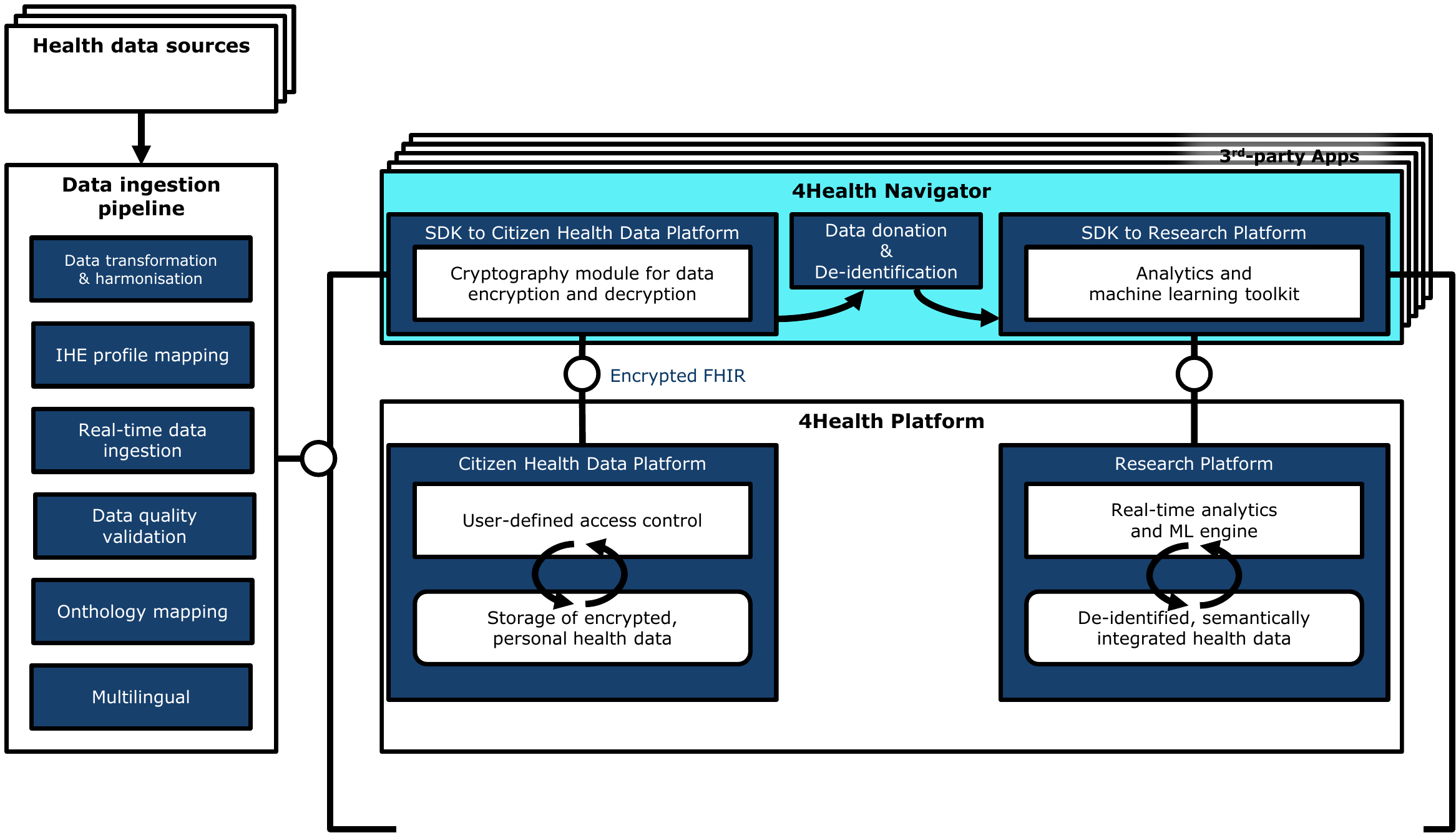}
	\caption{System architecture of the S4H software application.}
	\label{fig:s4h_architecture}
\end{figure*}

This section explains relevant aspects of the S4H software application that are needed to assess the feasibility of the integration of CEF DSIs. \autoref{fig:s4h_architecture} shows the system architecture.
The application is divided into the \newterm{4Health Platform} (4HP) that stores medical data, and the \newterm{4Health Navigator} (4HN), the citizen-facing front-end.

The 4HP contains the \newterm{Citizen Health Data Platform} (CHDP) and the \newterm{Research Platform} (RP). The CHDP holds all citizen data in encrypted form, which only the citizen can access; not even the platform provider can decrypt it. This so-called zero knowledge paradigm ensures data confidentiality and security, however, it introduces difficulties with regards to the implementation, since all operations that require to read data need to happen on client-side in the 4HN or 3-rd party apps.
When citizens donate their health data for research purposes, it is de-identified and stored in the RP, from which approved researchers can access it.

The 4HN contains basic functionalities for citizens, such as viewing, managing, sharing, and donating their health data. The connection between the 4HN and the 4HP is established by \newterm{software development kits} (SDKs) that take care of the authorized en- and decryption. The data format in which health data is stored and transferred is the Fast Healthcare Interoperability Resources (FHIR) standard.
3-rd party apps can also connect to the 4HP via SDKs, if they were registered by the platform provider and granted access by the user.

Different health data sources connect to the 4HP through the ingestion pipeline that transforms data to FHIR and can perform additional steps, such as data harmonization, quality validation, and ontology mapping.

\section{CEF DSIs}
\label{sec:cef_dsis}

This section describes CEF DSIs that qualify for the inclusion into the S4H software application: eID, eSignature \& eSeal, eTranslation, eDelivery, and eHealth.
Each DSI section \textit{3.X} is segmented into several subsections.
The CEF implementation in its current status is described in \textit{3.X.1}, use cases in S4H \textit{3.X.2}, and possibilities for the implementation in S4H \textit{3.X.3}.
	\subsection{eID}

According to the eIDAS Regulation, \newterm{electronic identification} is ``the process of using person identification data in electronic form uniquely representing either a natural person, or a natural person representing a legal person" (Art. 3(1)).
In contrast, \newterm{authentication} is defined as ``an electronic process that enables the electronic identification of a natural or legal person, or the origin and integrity of data in electronic form to be confirmed" (Art. 3(5)).
In order to authenticate, a person requires \newterm{electronic identification means}, ``a material and/or immaterial unit containing person identification data" (Art. 3(2)), that are issued under an \newterm{electronic identification scheme}, which ``means a system for electronic identification" (Art. 3(4)).

Moreover, the eIDAS Regulation specifies \textbf{assurance levels} for electronic identification schemes, namely \textit{low}, \textit{substantial}, and \textit{high}, referring to the ``degree of confidence in the claimed or asserted identity of a person" (Art. 8(2)).
To be recognized by other Member States in cross-border identification, the assurance level needs to be substantial or high. Additionally, to ensure interoperability, national electronic identification schemes need to comply to certain criteria, which are monitored in the \textbf{notification} process (Art. 9).
Different notification statuses exist~\autocite{eIDNotificationWebpage}:
\begin{enumerate}
	\item \textbf{Pre-notification} A national electronic identification scheme is pre-notified by transmitting relevant documents to the European Commission (EC) and other Member States.
	\item \textbf{Peer review} The electronic identification scheme is reviewed and assessed with regards to its compliance to technical specifications and level of assurance, which takes 3 months at the most.
	\item \textbf{Notification} The scheme is notified at the earliest 6 months after pre-notification.
	\item \textbf{Publication in OJEU} The notification is published in the \textit{Official Journal of the European Union} at the latest 2 months later.
	\item \textbf{Recognition by all Member States} Notified electronic identification schemes need to be recognized by other Member States 12 months after they were published in the OJEU.
\end{enumerate}
		\subsubsection{CEF Implementation}

The exchange of identification information is enabled by the decentralized \newterm{eIDAS Network} that consists of \newterm{eIDAS Nodes}~\autocite{eIDOverviewWebpage}.

This section comprises how the eIDAS Network and Nodes are defined, which national electronic identification schemes exist, which attributes are included, and how the eID solution was assessed so far.

\paragraph{eIDAS Nodes}

\begin{figure}[ht]
	\centering
	\includegraphics[width=0.25\linewidth]{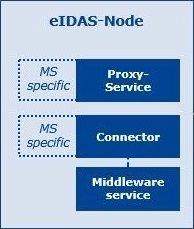}
	\caption{Key components of an eIDAS Node. Each node consists of a Member State specific Connector and a Proxy- or Middleware-Service (adapted from~\textcite{eIDDetailsWebpage}).}
	\label{fig:eid_components}
\end{figure}

A node is able to handle authentication requests and to translate between a national scheme and the common eIDAS format (see the network communication section below).
Nodes are maintained by the Member States.
\autoref{fig:eid_components} shows the main components, a \newterm{Connector} that is able to send identification requests to other Member States, and a \newterm{Proxy-} or \newterm{Middleware-Service} to receive them.
A Member State is either Proxy- or Middleware-based, while a Middleware service forwards Connector requests and therefore is not only deployed on the node of the receiving Member State but also on the sending one.

\paragraph{eIDAS Network Communication}

\begin{figure*}[ht]
	\includegraphics[width=\textwidth]{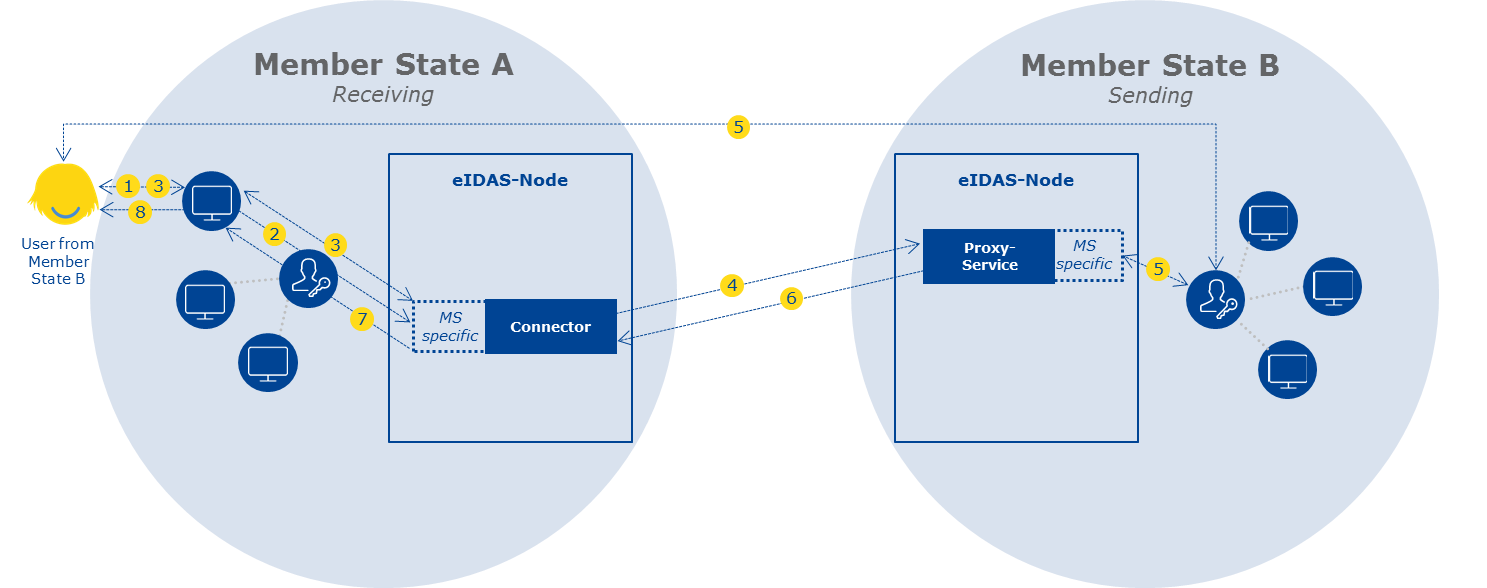}
	\caption{Communication between eIDAS Nodes of two proxy-based Member States (from~\textcite{eIDDetailsWebpage}).}
	\label{fig:eid_communication}
\end{figure*}

A schema of the communication~\autocite{eIDDetailsWebpage} between two countries is depicted in \autoref{fig:eid_communication}.
A citizen from Member State B wants to access a web service offered by a \newterm{Service Provider} in Member State A, for which he or she needs to authenticate (1).
The Service Provider sends a request for authentication to the \newterm{National Identity Provider} or directly to the Connector of Member State A (2).
When receiving the request receipt, the Service Provider submits the state of origin as indicated by the citizen, in this case Member State B (3).
The Connector now creates an \newterm{eIDAS Request} (4), which is translated to a request for the National Identity Provider of Member State B and enables the citizen to authenticate against the regarding national eID (5).
The authentication result is transferred in an \newterm{eIDAS Response} to the eIDAS Node of Member State B and sent back to the Connector of Member State A (6), where a response for the Service Provider is created and transferred either directly or via the National Identity Provider of Member State A (7).
If the authentication was successful, the Service Provider grants access to the citizen (8).

Here, we only describe the case of two proxy-based countries; for middleware-based communication the exact procedures differ, however, this merely effects the communication between nodes and does not influence the interfaces for citizens and Service Providers.

\paragraph{Notification}

As stated before, the eIDAS Regulation demands the recognition of notified eID schemes one year after their publication in the OJEU.
\autoref{tab:notification} shows an overview of the notification statuses of Member States that are part of the S4H consortium.
Some Member States have notified multiple electronic identification schemes; here, the overall status is determined by the scheme with the highest status (notified $>$ pre-notified $>$ peer-reviewed).
\begin{table}[ht]
	\caption{Member States participating in the S4H consortium with their status in the notification process, ordered by the date of publication in the OJEU if notified (N), or the peer-review (R). Member States with status other (O) have not notified their electronic identification scheme~\autocite{notificationOverview}.}
	\label{tab:notification}
	\newcommand{\status}[2]{{\color{#1}#2}}
	\newcommand{\notified}{\status{NavyBlue}{N}}
	\newcommand{\prenotified}{\status{Dandelion}{P}}
	\newcommand{\peerreviewed}{\status{ForestGreen}{R}}
	\newcommand{\other}{\status{gray}{O}}
	\centering
	\begin{tabular}{rlr}
		\toprule
		\textbf{Member State} & \textbf{Status} & \textbf{Date}\\
		\midrule
		Germany & \notified & 26 Sep 2017\\
		\midrule
		Italy & \notified & 10 Sep 2018\\
		\midrule
		Luxembourg & \notified & 07 Nov 2018\\
		\midrule
		Belgium & \notified & 27 Dec 2018\\
		\midrule
		Portugal & \notified & 28 Feb 2019\\
		\midrule
		The Netherlands & \peerreviewed & 06 Jun 2019\\
		\midrule
		Austria & \other & \textit{In use}\\
		\midrule
		France & \other & \textit{In development}\\
		\bottomrule
	\end{tabular}
\end{table}
The support of not notified schemes is voluntary.

\paragraph{Minimum Data Set (MDS)}

The Commission Implementing Regulation (EU) 2015/\-1501 defines a ``minimum data set of person identification data uniquely representing a natural or legal person, pursuant to Article 12(8) of the eIDAS Regulation"~\autocite{eu2015/1501}.

For a natural person, the mandatory attributes are):
\begin{itemize}
	\item Current family name(s)
	\item Current first name(s)
	\item Date of birth
	\item Unique identifier
\end{itemize}
The unique identifier is determined as ``constructed by the sending Member State in accordance with the technical specifications for the purposes of cross-border identification and which is as persistent as possible in time"~\autocite{eu2015/1501}.
This is implemented differently by Member States, Germany for example creates a pseudonym that depends on the requesting Member State (for public-sector bodies) or the relying party (for non-public-sector bodies)~\autocite{eIDGermany}; other countries use tax codes or national personal identification codes~\autocite{attributesOverview}.

Additional attributes defined in the MDS are:
\begin{itemize}
	\item First name(s) and family name(s) at birth
	\item Place of birth
	\item Current address
	\item Gender
\end{itemize}
Moreover, Member States can provide further attributes voluntarily.

\autoref{tab:attributes} lists which attributes are contained in the MDS of electronic identification schemes of Member States participating in the S4H consortium.
Again, for multiple schemes per Member State, the one with the highest status is included.
\begin{table}[ht]
	\caption{Attributes of the MDS contained in electronic identification schemes of Member States that are part of the S4H consortium~\autocite{attributesOverview}. For multiple schemes per Member State the one with the highest status is included. LN: current family name(s), FN: Current first name(s), BD: Date of birth, ID: Unique identifier, BN: First name(s) and family name(s) at birth, BP: Place of birth, A: Current address, G: Gender.}
	\label{tab:attributes}
	\centering
	\newcommand{\present}{\greencheck}
	\newcommand{\missing}{\redcross}
	\newcommand{\unknown}{\grayquestionmark}
	\begin{tabular}{rcccccccc}
		\toprule
		\textbf{\linebreakcell{Member State}} &
		\textbf{\linebreakcell{LN}} &
		\textbf{\linebreakcell{FN}} &
		\textbf{\linebreakcell{BD}} &
		\textbf{\linebreakcell{ID}} &
		\textbf{\linebreakcell{BN}} &
		\textbf{\linebreakcell{BP}} &
		\textbf{\linebreakcell{A}} &
		\textbf{\linebreakcell{G}}\\
		\midrule
		Germany & \present & \present & \present & \present & \present & \present & \present & \missing \\
		\midrule
		Italy & \present & \present & \present & \present & \present & \missing & \missing & \missing \\
		\midrule
		Luxembourg & \present & \present & \present & \present & \missing & \present & \present & \present \\
		\midrule
		Belgium & \present & \present & \present & \present & \missing & \present & \missing & \present \\
		\midrule
		Portugal & \present & \present & \present & \present & \missing & \missing & \present & \present \\
		\midrule
		The Netherlands & \present & \present & \present & \present & \missing & \present & \present & \missing \\
		\midrule
		Austria & \unknown & \unknown & \unknown & \unknown & \unknown & \unknown & \unknown & \unknown \\
		\midrule
		France & \unknown & \unknown & \unknown & \unknown & \unknown & \unknown & \unknown & \unknown \\
		\bottomrule
	\end{tabular}
\end{table}

\paragraph{Assessment}

The basic architecture was presented by Carretero et al.~\autocite{carretero2018}; they assess the European eID solution as performant and scalable.

Engelbertz at al.~\autocite{engelbertz2018} were able to identify vulnerabilities in the eIDAS Node reference implementation that enable Denial-of-Service (DoS) and Server Side Request Forgery (SSRF) attacks.
They propose best-practices and provide a test-suite to support developers of eID services in improving the security even further.

Berbecaru et al.~\autocite{berbecaru2019} show how the eIDAS attribute set can be extended. They implement an additional attribute provider, using the example of the Erasmus exchange between universities.


\subsubsection{Use Cases}

For all use cases additional information is provided: Those that do not originate directly from the grant agreement are marked with an asterisk (\textasteriskcentered). Use cases considered applicable are preceded by a check mark (\greencheck{}) and not applicable by an x mark (\redcross{}). Cases where open questions remain, are marked with a question mark (\grayquestionmark{}).

First of all, the eID building block could be used for the validation of a citizen to ensure that the given name and possibly other information are connected to a real person. From this, different use cases emerge:

\grayquestionmark{} \textbf{Trust in sharing}

\textit{Description:} In order to share medical content with each other, users might want to be sure that the user they grant access to is the person he or she pretends to be.

\textit{Assessment:} A validation would probably add trust in sharing.
However, using the handshake sharing as it is currently envisioned to be implemented, the user sharing his or her data and the grantee are in direct communication to exchange a PIN.
If this is done in person, as for example when sharing content with a doctor in an appointment, the trust is established by being face to face, an additional validation would be redundant.
If the appointment is happening remotely, such an additional validation could be more valuable.
A third scenario is an asynchronous remote sharing, for example preceding an appointment as preparation for the health care professional (HCP).
Here, a validation of the sharing party is crucial.
In general it shall be noted that purely relying on the digital identification of the user introduces a high risk of abuse due to identity theft, therefore, additional authentication factors should be considered.

\textit{Open questions:} Is this functionality desired by the citizen?

\grayquestionmark{} \textbf{Part of eSignature service}

\textit{Description:} eID can be used as one part of the e\-Signature service to enable an advanced or qualified signature (see \autoref{sec:esignature}).

\textit{Assessment:} This highly depends on whether a qualified signature is needed and how it will be implemented.

\grayquestionmark{} \textbf{Check prerequisites for platform usage*}

\textit{Description:} The S4H software application shall---at least in its initial testing phase---only be accessible to EU citizens of legal age, which could be ensured by using eID.

\textit{Assessment:} It might be legally needed to verify usage conditions.
However, it remains questionable whether the EU citizenship can be proven, since nationality is not accessible for most national electronic identification schemes.
Moreover, for some Member States also enable electronic identification with electronic residence permits; if the usage condition is extended to long-term residents, this approach is actionable.

\textit{Open questions:} Is it legally needed to verify usage conditions?

\redcross{} \textbf{Legal fallback for platform provider*}

\textit{Description:} Technically, not only medical data can be uploaded and shared by users to the CHDP, but every kind of data. To ensure a legal fallback in case of illegal content, verified users could be held accountable for their uploads.

\textit{Assessment:} Since the platform provider can neither access uploaded content nor detailed user information due to the zero knowledge paradigm, an identification of inappropriate content and the real name of the user who uploaded it is not possible, at least from the provider perspective.

Additional use cases are presented as follows:

\grayquestionmark{} \textbf{Use as authentication factor}

\textit{Description:} eID could be used as an authentication factor for log-in.

\textit{Assessment:} Using eID would probably be inconvenient compared to other well-established options, such as using email and phone.

\textit{Open questions:} Is using eID as an authentication factor convenient and desired by the user?

\redcross{} \textbf{Connect to other systems using eID*}

\textit{Description:} One problem is to map users in one system to users in another; this could be approached using eID.

\textit{Assessment:} eID is not usable to map users between systems for all national electronic identification schemes, as the unique identifier can be Member State or service dependent.
When HCPs share data with users, it needs to be validated that the receiving party is the correct person; in the ingestion context this is solved similar to handshake sharing, using email and phone number.
Moreover, the citizen usually is present when the connection is established, which does not require further identification.
If this is done remotely, an additional factor to ascertain a user identity might be useful.
However, according to current knowledge the identification will more likely be covered by email, account name, phone number or similar identifiers that make sure the destination the document is sent to belongs to the right patient.

\textit{Open questions:} What are the needs and the workflow of the hospital or doctor with regards to sharing?

\redcross{} \textbf{Use for registration}

\textit{Description:} eID could be used for the registration of a citizen.

\textit{Assessment:} Across all Member States, only few attributes are required, namely the first name, family name, date of birth, and an unique identifier.
For a proper registration, additional attributes such as the current address, e-mail address, and phone number would be needed.
However, if a user wishes to validate his or her account while registering, the present fields could be filled automatically.

		\subsubsection{Implementation in S4H}

To access the eIDAS Network, the connection to one eIDAS Node is needed. In our case, the German eID node is the most reasonable, since the 4HP is planned to be hosted in Germany. It requires four steps to gain access:

\begin{enumerate}
	\item \textbf{Conceptualize the service and which attributes are needed} The first step is to conceptualize the service---the S4H software application---to determine, which attributes are needed and why.
	As shown in \autoref{tab:attributes}, the national eID schemes support different sets of attributes, so two possibilities exist here:
	Either include as much attributes as possible and deal with missing values, or only include the four mandatory attributes.
	The latter case seems more reasonable in order to create a common look-and-feel across all Member States and to keep the implementation of functionalities involving eID less complex.
	Moreover, it needs to be considered that the same attribute can be implemented in different ways, as for example the unique identifier that can be static or service dependent.
	
	\item \textbf{Request authorization} To connect to the German eID service, authorization needs to be requested from the Federal Office of Administration, in written form or personally~\autocite{becomeSP}. The authorization is only valid for a specific amount of time and can be extended.
	
	\item \textbf{Choose certificate and service providers}  Once authorized, a provider for an authorization certificate needs to be chosen, as well as an eID service provider; alternatively, an own eID server can be set up and certified. Since the implementation of an own eID server requires a higher effort and currently no consortium partner plans to become an eID service provider itself, the connection to an existing eID service is more practical.
	
	\item \textbf{Connect services} Finally, the S4H software application needs to be connected to the service or server. How the connection between application and identification service is established in detail depends on the specific provider or server implementation. Furthermore, the set of supported electronic identification schemes can differ---apart from the mandatory set of notified schemes. Since user interaction is required, the eID connection needs to be established via the 4HN.
\end{enumerate}

	\subsection{eSignature \& eSeal}
\label{sec:esignature}

In the eIDAS Regulation, three types of electronic signatures are defined.

A basic \newterm{electronic signature} is ``data in electronic form which is attached to or logically associated with other data in electronic form and which is used by the signatory to sign" (Article 3(10)), while the \newterm{signatory} is a natural person (Article 3(9)).
This means, an electronic signature can simply be a name typed under an email or a checkbox clicked by a user.

\begin{figure}[ht]
	\centering
	\includegraphics[width=0.8\linewidth]{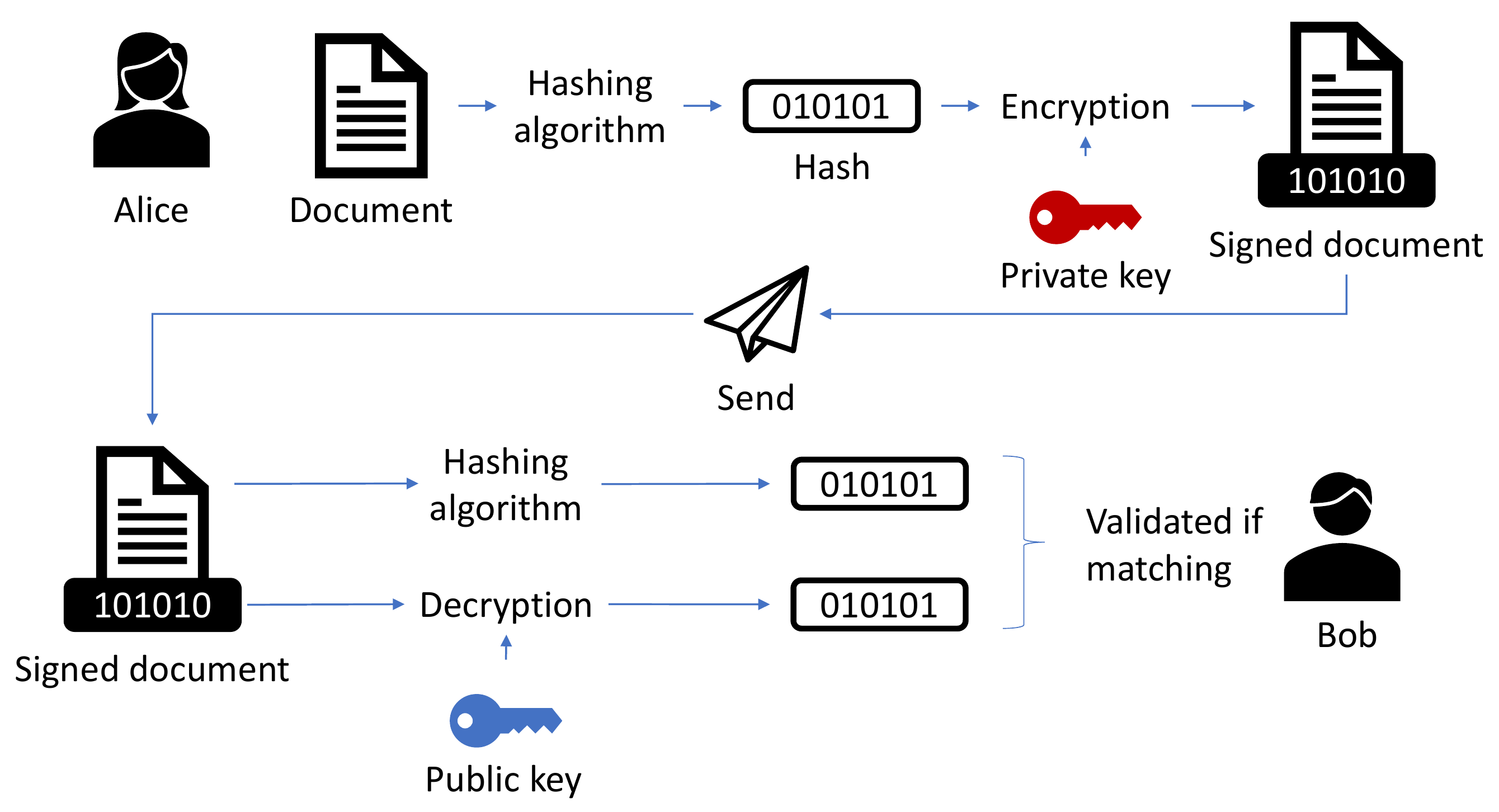}
	\caption{The creation and validation of a digital signature; Alice wants to sign a document and send it to Bob. The data is signed by creating a hash, encrypting it with Alice's private key, and attaching it to the document. When Bob receives the signed data, he creates the hash himself and decrypts the signature using Alice's public key, which usually is stored in a certificate. If both hashes match the signature is valid; it means that Alice is the signer and the document was not changed (adapted from~\textcite{digitalSignature}).}
	\label{fig:signature}
\end{figure}

An \newterm{advanced electronic signature} `is uniquely linked to the signatory, capable of identifying the signatory created using electronic signature creation data that the signatory can, with a high level of confidence, use under his sole control, and linked to the data signed therewith in such a way that any subsequent change in the data is detectable' (Article 26). This for example can be accomplished using a digital signature based on a certificate (see \autoref{fig:signature} for details).

The most powerful type of electronic signature is the \newterm{qualified electronic signature} that has ``the equivalent legal effect of a handwritten signature" (Article 25(2)).
It is ``an advanced electronic signature that is created by a qualified electronic signature creation device, and which is based on a qualified certificate for electronic signatures" (Article 3(12)).
A \newterm{certificate for electronic signatures} is ``an electronic attestation which links electronic signature validation data to a natural person and confirms at least the name or the pseudonym of that person" (Article 3(14)); a \newterm{qualified certificate for electronic signatures} is ``a certificate for electronic signatures, that is issued by a qualified trust service provider"(Article 3(15)) and moreover meets certain criteria laid down in the eIDAS Regulation, such as that it contains ``details of the beginning and end of the certificate’s period of validity" (Annex I).
An \newterm{electronic signature creation device} is ``configured software or hardware used to create an electronic signature".
To become a \newterm{qualified electronic signature creation device}, it has to meet further requirements, such as to ensure that ``the confidentiality of the electronic signature creation data used for electronic signature creation is reasonably assured" (Annex II).

The German eID smart card, for example, is an electronic signature creation device that holds a qualified certificate for electronic signatures and is capable to create qualified signatures.

\textbf{Electronic seals} follow similar regulations as electronic signatures but for legal instead of natural persons. Primarily, the type of certificate needed for a qualified electronic seal differs.

		\subsubsection{CEF Implementation}
\label{sec:esignature_cef_bb}

A sample implementation for the eSignature building block is given in form of the open-source, Java-based \newterm{Digital Signature Service} (DSS) that is able to sign documents and validate signatures based on various different standards~\autocite{dssDocs}.
The procedure to create an advanced or qualified electronic signature is described as follows~\autocite{eSignatureFAQ}:

\begin{enumerate}
	\item \textbf{Obtain certificate} The first step is to obtain a digital certificate from a trust service provider; for a qualified signature, it needs to be a qualified certificate. (Qualified) trust service providers per country can be found on the European List of Trusted Lists via the Trusted List Browser\footnote{\url{https://webgate.ec.europa.eu/tl-browser/\#/}}.
	\item \textbf{Sign document} In a second step, the document is signed; a service to do this might be offered by the Trusted Service Provider. Additionally, the DSS demonstration app\footnote{\url{https://ec.europa.eu/cefdigital/DSS/webapp-demo/sign-a-document}} can be used; it includes the open-source NexU browser signing application\footnote{\url{http://nowina.lu/solutions/java-less-browser-signing-nexu/}} that supports different European authentication and eSignature devices.
\end{enumerate}

It shall be noted that Engelbertz at al.~\autocite{engelbertz2019} found ``several vulnerabilities" regarding the DSS validation of  XML Advanced Electronic Signatures (XAdES). They were able to `read server files and bypass XAdES protections' using DoS, SSRF, and XML Signature Wrapping (XSW) attacks.
		\subsubsection{Use Cases}

\grayquestionmark{} \textbf{Sign consent}

\textit{Description:} In S4H, different types of consents exist, e.g., for platform usage and data donation. The user needs to accept the consent and potentially sign it using eSignature.

\textit{Assessment:} It might be legally needed to sign a consent with a legally binding (qualified) signature.

\textit{Open questions:} Is it legally needed to sign a consent with a legally binding signature?

\grayquestionmark{} \textbf{Use as proof of origin for third parties in ingestion*}

\textit{Description:} If for example a hospital (providing HCP) shares a medical document, it can sign it digitally using eSeal to make sure that (1) the medical document was not changed after signing and (2) it was issued by this specific hospital. This could be especially reasonable if the citizen that receives the document shares it with another HCP, for example his or her general practitioner (consuming HCP); the consuming HCP also might want to ensure the integrity and origin of the document.

\textit{Assessment:} By using a registered SDK for ingestion and encryption, the integrity and origin already is ensured by the platform. However, it might be legally needed to validate it by an official third party.

\textit{Open questions:} Is the validation of integrity and origin by an official third party legally needed for the providing or the consuming HCP? Are HCPs capable to provide qualified electronic seals by themselves or is an implementation in S4H needed?

\grayquestionmark{} \textbf{Use as proof of origin for professional users that are not part of S4H*}

\textit{Description:} HCPs that are not part of S4H, which means they are not connected through the ingestion pipeline, shall after all be able to upload content for the user. To ensure data integrity and proof the origin of uploaded content, a digital signature could be used.

\textit{Assessment:} This use case was not largely considered yet; however, such a mechanism is useful.

\textit{Open questions:} Is the validation of integrity and origin by an official third party legally needed for the providing or the consuming HCP?
		\subsubsection{Implementation in S4H}

This section lists different possibilities of the implementation of the eSignature and eSeal CEF building block in the S4H software application.
It first describes the signature and seal creation before it refers to the validation of eSignatures and eSeals.

The type of electronic signature which is here referred to is the qualifying signature, as it yields the highest complexity and implicates the other types.

It shall be noted that for the implementation of a basic or an advanced electronic signature or seal no CEF building block is required. However, an advanced electronic signature or seal \textit{can} be created using CEF building blocks; the difference to the qualified case is that certificates and creation devices do not need to be qualified.

\paragraph{eSignature Creation}

To create a qualified electronic signature, a qualified certificate for electronic signatures and a qualified electronic signature creation device are needed.
This introduces difficulties, since various formats of certificates and creation devices exist in different Member States that follow several standards and provide specific interfaces, as for example smart cards.
In the following, implementation scenarios that enable qualified signatures are presented.

\textit{(1) No signature service}

Since users need to obtain a qualified certificate first, the easiest solution---from the implementation perspective---is to pass the responsibility of signing documents to the user and only include validation capabilities in the S4H software application. The signing, once a certificate was obtained, can for example simply be done using Adobe Reader or the DSS demonstration app. However, this is the least convenient solution for the user.

\textit{(2) Use DSS}

The DSS sample implementation could be installed on-premises as a web service on the 4HP that can be accessed by the 4HN.
The zero knowledge paradigm is not considered impaired, since the use case for electronic signatures only involves signing consents that do not hold person-specific data and need to be accessible by the platform anyway.
However, this approach would require to implement interfaces for various Member State specific types of authentication and signature devices or to rely on third-party solutions, such as NexU (see \autoref{sec:esignature_cef_bb}).
The usability of this approach highly depends on the signature creation device, whether it already comes with a qualified signature, and how well the integration into the DSS and S4H software application is implemented.

\textit{(3) Use remote service}

As an alternative to an on-premises installation that requires the additional implementation of interfaces and maintenance, a third-party service for qualified eSignatures could be used. Such a service should be able to grant and manage qualified certificates, and create qualified signatures for documents it receives. This merely requires the connection to the regarding service; however, data would be shared outside of the S4H context and again, the convenience does highly depend on the remote service.

\textit{(4) Become a trust service provider}

To avoid that data is passed to third-parties while having a unified way to acquire qualified certificates and create qualified signatures, the S4H software application or one member of the consortium would need to become a qualified trust service provider themselves.
Although building blocks such as eID and the DSS could be re-used, this poses the highest implementation effort.
Additionally, an administrative burden emerges due to the required accreditation by one Member State, for example in Germany by the Federal Network Agency~\autocite{getQualified}.
However, the usability would only depend on the implementation.

Option (3) yields the best trade-off between implementation effort and usability; however, it remains unclear whether the integration of an eIDAS complying service other than the DSS meets the requirements of an implementation of the eSignature CEF building block as stated in the grant agreement.
Moreover, further investigation is needed in order to identify suitable services.
Option (2) states the second-best solution.
Due to low implementation costs option (1) seems tempting, however, the S4H software application would possibly loose users due to inconvenience.
If electronic signatures only need to be applied in a few not mandatory occasions, this option could remain feasible.

\paragraph{eSeal Creation}

As for a qualified electronic signature, for a qualified electronic seal a qualified certificate and a qualified creation device are needed.
With the side node that HCPs usually do not possess smart cards that enable them to create qualified electronic seals and therefore need to apply for such a certificate, the above mentioned options also apply here.
Moreover, some institutions might already be able to create qualified electronic seals, which would make option (1) more feasible.
In either case, eSeal raises the complexity for HCPs to connect to the S4H software application.

\paragraph{Validation}

Different than for the signature creation, for validation purposes the DSS can be included more easily into the S4H software application, since various types of signatures are supported.
To validate an electronic signature or seal, the data needs to be present in decrypted form.
For eSignature, an on-premises installation of the DSS could be used.
For eSeal the validation could happen as one step of the ingestion pipeline, possibly also using the on-premises DSS web service.
However, to validate the origin of a sealed document, e.g. by a third party, further steps are needed, since certificates are only valid for a certain amount of time.
One possibility is to store the certificate the eSeal was created with, which requires additional storage and its implementation.
Moreover, it is possible to extend an electronic seal or signature using a (qualified) timestamp once the certificate expires; but, although lasting longer, the timestamp certificates also expire and need to be extended.
This generates efforts on the implementation side and, additionally, for the third party that attempts to validate a document's origin.
The question here is, how qualified such a validation of origin must be and how long it needs to be possible (if required at all).

	\subsection{eTranslation}

The eTranslation building block is the only CEF building block mentioned in this report that is not described by the eIDAS Regulation. It is a translation service building ``on the previous machine translation service of the European Commission, MT@EC, developed by the Directorate-General for Translation (DGT)"~\autocite{eTranslationOverview}.
		\subsubsection{CEF Implementation}

The CEF eTranslation building block is a translation service based on neural machine translation~\autocite{eTranslationDocs}.
It was trained on the Euramis translation memories that include over 1 billion sentences in the 24 official EU languages, most originating from legislative documents; therefore, eTranslation can translate between all these languages and is ``particularly suited for the needs of EU policy documents".
As an input, either plain text or formatted files in various formats can be provided.
The service is accessible in two ways, either via a web-interface for human-to-machine use, or through an API for machine-to-machine use.

To further improve the quality of translations in different domains and languages, `the European Commission has launched a comprehensive European Language Resource Coordination (ELRC) effort to identify and gather relevant language and translation data'. 
		\subsubsection{Use Cases}

\greencheck{} \textbf{Translate medical content}

\textit{Description:} In the cross-border case, e.g., when emergencies occur on vacation, at least basic and possibly most recent medical information needs to be translated to a language that the attending physician can understand.

\textit{Assessment:} For a pan-European usage, the translation of medical content is definitely needed.

\textit{Open questions:} Are eTranslation results suitable for medical language? This also applies to medication---not the name but rather the active ingredient should be translated. Possibly, the eHDSI terminology services (see \autoref{sec:ehealth}) could be included for this purpose.

\redcross{} \textbf{Translation of medical codes and user interface}

\textit{Description:} eTranslation could be used for the translation of static content, such as medical codes and text in the user interface.

\textit{Assessment:} The translation of static content is usually solved manually, however, eTranslation could be applied here, too. For codes from medical coding systems such as ICD-10, LOINC, and SNOMED-CT, rather coding system specific utilites should to be applied than a general translation service.

Additionally, the grant agreement states that training and education material should be available in several EU languages; this is not a central part of the S4H software application, wherefore it is not examined further in this report.
		\subsubsection{Implementation in S4H}

For the implementation of the eTranslation CEF building block two major questions emerge: First, whether the S4H application is allowed to use the CEF building block in the first place, including its usage after the project ended, and second, where to include eTranslation.

\paragraph{Usage Conditions}

The official eTranslation documentation states that ``European and national public administrations, or cross-border EU projects supported from the CEF programme can integrate eTranslation in their digital services"~\autocite{eTranslationIntegration}. For the S4H project it remains unclear whether it is allowed to use eTranslation at all since it is not funded directly under the CEF programme. If it is, it needs to be clarified whether the CEF building block can still be used in the S4H software application after the project ends.

\paragraph{Integration into S4H}

Similarly to the integration of eSignature, different possibilities exist to include a translation service into the S4H software application architecture.

\textit{(1) Ingestion pipeline}

The translation component could be included in the ingestion pipeline, so that each document is stored in all available languages; however, this is not a feasible approach, since it does neither support the improvement of translation algorithms over time nor the extension of available languages.

\textit{(2) CHDP}

For an integration in the CHDP, the encrypted user content needs to be decrypted, which impairs the zero knowledge paradigm and therefore is only feasible if the encryption is softened vigorously.

\textit{(3) On-premises installation}

Another possibility to include translation as a back-end service is to host it on-premises as a web service that can be accessed by the 4HN, where decrypted data is accessible. Furthermore, the user could give his or her consent that the translation service is allowed to read clear text data. In context of the eTranslation CEF building block another question emerges, namely, whether it even is possible to host it on-premises.

\textit{(4) API-based}

As an alternative, a third-party remote translation service, such as the eTranslation API, could be included. However, in this case the decrypted user data is passed to third-parties; with the above mentioned on-premises approach we could at least guarantee that the data does not leave the S4H context.

Additionally, both the on-premises and the third-party translation could be combined with a de-identification service. This indeed increases data security but probably decreases the translation quality and thereby usability. Moreover, it introduces additional automated steps that may reduce data quality. More importantly, the de-identification service also needs to read decrypted data, which introduces the same problems as mentioned for eTranslation.

In summary, an on-premises or API-based approach without de-identification seems most reasonable. However, the data confidentiality could be impaired in case the services illicitly store or analyze data.

	\subsection{eDelivery}

The eIDAS Regulation defines an electronic registered delivery service (ERDS) as ``a service that makes it possible to transmit data between third parties by electronic means and provides evidence relating to the handling of the transmitted data, including proof of sending and receiving the data, and that protects transmitted data against the risk of loss, theft, damage or any unauthorised alterations" (Art. 3(36)).
As for eSignature and eSeal, ``data sent and received using an electronic registered delivery service shall not be denied legal effect and admissibility as evidence in legal proceedings solely on the grounds that it is in an electronic form or that it does not meet the requirements of the qualified electronic registered delivery service" (Art. 43).
However, for a qualified ERDS service additional properties can be presumed: The identification of sender and addressee is ensured, as well as the data integrity, by using ``an advanced electronic signature or an advanced electronic seal of a qualified trust service provider" (Art. 44(1d)). Additionally, ``any change of the data needed for the purpose of sending or receiving the data is clearly indicated to the sender and addressee of the data" (Art. 44 (1e)).
Lastly, ``the date and time of sending, receiving and any change of data are indicated by a qualified electronic time stamp" (Art. 44(1f)).
		\subsubsection{CEF Implementation}

\begin{figure*}[ht]
	\includegraphics[width=\textwidth]{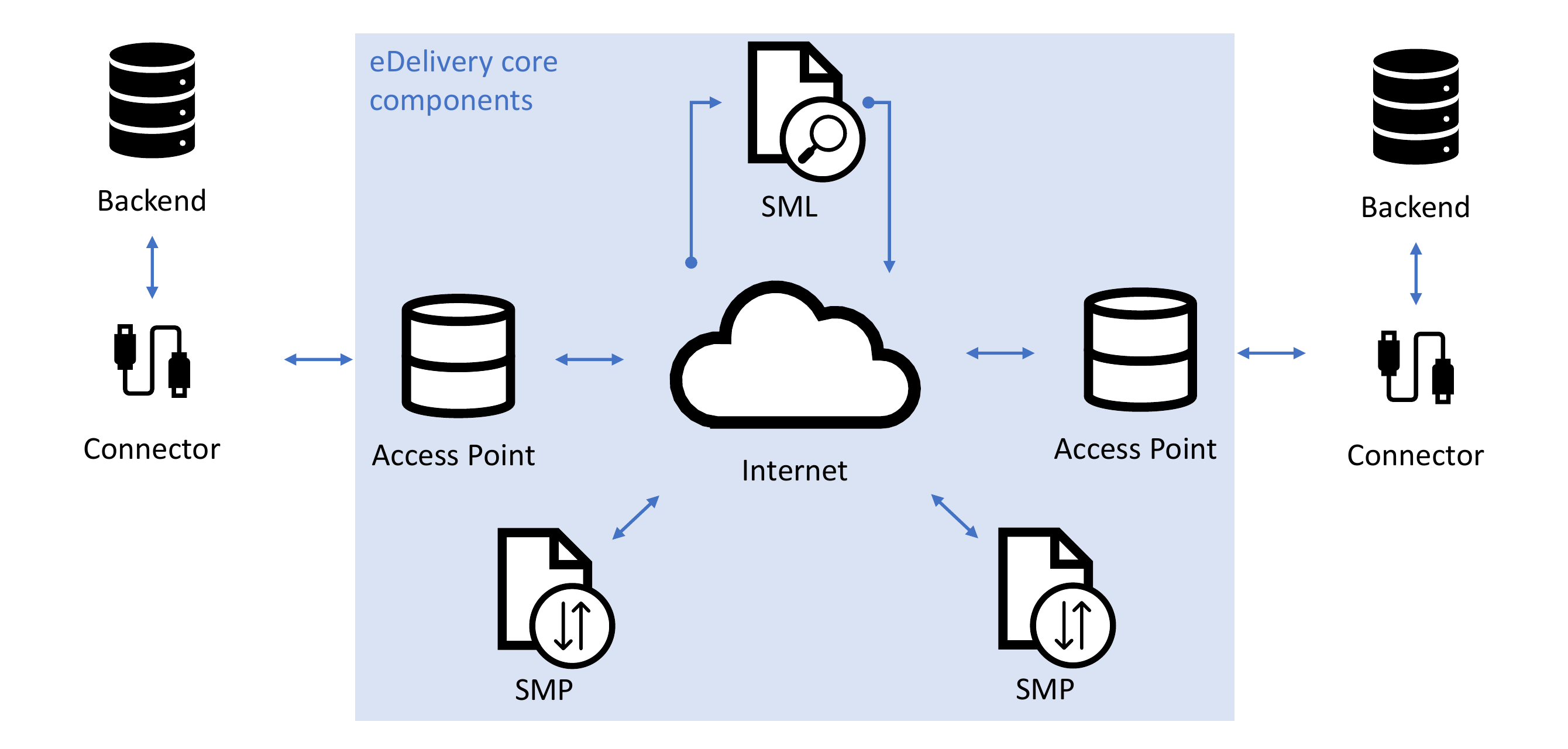}
	\caption{The elements of eDelivery. The back-ends of two parties are connected to Access Points that interact with each other. Dynamic service and capability lookup is enabled by SMP (Service Metadata Publisher) and SML (Service Metadata Locator) components (adapted from~\textcite{eDeliveryDocs}).}
	\label{fig:edelivery}
\end{figure*}

The CEF eDelivery building block defines technical requirements for a secure and reliable data exchange between different parties, that works independently from the particular technical environment~\autocite{eDeliveryDocs}.
It is based on encryption and digital signatures, and provides legal assurance that data is delivered once (and only once), even when the receiving party is temporarily unavailable.
The overall architecture of the eDelivery solution is displayed in \autoref{fig:edelivery}.

The building block is structured in a so-called \newterm{4-corner model}, where the communicating systems do not interact with each other directly but via \newterm{Access Points} that exchange data based on specific standardized protocols, namely ``the AS4 Profile developed by e-SENS or of the AS2 Profile developed by OpenPEPPOL"~\autocite{accessPointDocs}.
Access Points can be integrated with the regarding back-ends via \newterm{Connectors}.

The \newterm{Service Metadata Publisher} (SMP) and \newterm{Service Metadata Locator} (SML) components enable a dynamic service location and capability lookup. The SMP holds metadata about every participant in the exchange network, such as the IP address of the Access Points, supported protocols, business processes, and the security setup, including certificates. The SML enables Access Points to locate the SMP of the network participant they want to contact.

For all components, sample implementations exist\footnote{Access Point with Connector plugins: \url{https://ec.europa.eu/cefdigital/wiki/display/CEFDIGITAL/Domibus}}\footnote{SMP: \url{https://ec.europa.eu/cefdigital/wiki/display/CEFDIGITAL/SMP}}\footnote{SML: \url{https://ec.europa.eu/cefdigital/wiki/display/CEFDIGITAL/SML+4.0.0}}. The software components are intended to be deployed as part of a back-end setup on an application server (Apache Tomcat or WebLogic) with a dedicated database (MySQL or Oracle 10g).

Additionally, the CEF eDelivery PKI (Public Key Infrastructure) service~\autocite{eDeliveryPKI} can be used to issue and manage digital certificates for signing and encrypting messages.

		\subsubsection{Use Cases}

\grayquestionmark{} \textbf{CHDP or RP to ELIXIR}

\textit{Description:} eDelivery could be used for secure data transfer of donated data to ELIXIR.

\textit{Assessment:} It is not clear yet, how data will be transferred and how secure the data transfer needs to be.

\textit{Open questions:} Is eDelivery suitable for the data transfer to ELIXIR?

\grayquestionmark{} \textbf{RP to researcher*}

\textit{Description:} eDelivery could be used to transfer data from the RP to the researcher.

\textit{Assessment:} As for the data transfer to ELIXIR, it is not clear yet, how---or whether---data will be transferred and how secure the data transfer needs to be. It needs to be discussed whether researchers can access the data directly or only are allowed to work on the RP.

\textit{Open questions:} Is eDelivery suitable for data transfer to researchers and will data be transferred after all?

\grayquestionmark{} \textbf{Communication with NCPs for eHealth*}

\textit{Description:} A connection between NCPs for eHealth (see \autoref{sec:ehealth}) and the CHDP or ingestion pipeline could be established using eDelivery, especially in case it is a regular feature of the eHealth DSI.

\textit{Assessment:} As in the before mentioned cases, it is not clear yet, how data will be transferred and how secure the data transfer needs to be. Moreover, whether eDelivery is a reasonable solution also depends on the architecture of NCPs.

\textit{Open questions:} Is eDelivery suitable for data transfer from NCPs?

\redcross{} \textbf{4HP to 4HN}

\textit{Description:} eDelivery could be used to securely transfer data from the 4HP to the 4HN.

\textit{Assessment:} The SDK already is conceptualized to take care of this functionality, while encrypting the data and establishing an additional end-to-end encrypted communication channel.

\redcross{} \textbf{Part of ingestion pipeline}

\textit{Description:} eDelivery could be used for the ingestion, to send data from a HCP to the 4HP.

\textit{Assessment:} A large part of the ingestion, including encryption and a secure data transfer, is handled by the SDK. eDelivery could be used to send data from the medical information system to the ingestion pipeline; whether this is practicable depends on whether the SDK is deployed inside the HCP network---a secured environment---or not.

\textit{Open questions:} Will the SDK be deployed inside the HCP network or will unencrypted data need to be transferred outside of this secure environment?
		\subsubsection{Implementation in S4H}

The CEF sample implementations for the Access Point, the SML, and the SMP require an application server and a database; therefore, they rather are suited to be included in the back-end, namely the 4HP or the ingestion pipeline, than the front-end, the 4HN. Moreover, the Access Point needs to be integrated with the regarding back-end systems. The parties that the 4HP communicates with also need to install an SMP and Access Point to be able to exchange information.

For the ingestion pipeline, eDelivery only states a reasonable choice if not all components can or shall be deployed in the secure environment of the HCP and therefore a reliable data transfer to a remote service is needed.

Additionally, for the communication between eDelivery components, certificates are needed. In case the eDelivery service shall or needs to be qualified, qualified timestamps need to be created, as well as at least advanced electronic signatures or seals issued by qualified trust service providers.

As an alternative to an own implementation, a suitable (qualified) trust service for ERDS represented in the European List of Trusted Lists could be used.
	\subsection{eHealth}
\label{sec:ehealth}

The eHealth DSI (eHDSI) is being developed in an ongoing EU project and aims to provide services and infrastructures that enable cross-border healthcare facilities~\autocite{eHealthOverview}.
It is defined by two main use cases that particularly matter abroad: The \newterm{Patient Summary} that holds key health data of a patient, which supports health professionals in unplanned care encounters, and the \newterm{ePrescription} that enables patients to receive their medication.

To ensure interoperability between countries, the eHDSI architecture consists of only few central services, including facilities for configuration and terminology. The focus lies on distributed services per Member State, in form of national contact points for eHealth (NCPeHs, in the following called NCPs).
NCPs of different countries are able to exchange health data, such as Patient Summaries and ePrescriptions, on top of national data formats and schemes in a standardized way.

The implementation part of the eHealth project is planned to finish in December 2020.
Recently, in January 2019, the ePrescription exchange between Finland and Estonia started, Finish patients can now receive their medication in Estonia~\autocite{eHealthArticle}.
10 Member States (Finland, Estonia, Czechia, Luxembourg, Portugal, Croatia, Malta, Cyprus, Greece and Belgium) may start these exchanges by the end of 2019.
In total, 22 Member States are part of the eHDSI and are expected to exchange ePrescriptions and Patient Summaries by the end of 2021.

The S4H project will not implement functionalities already incorporated in the eHDSI, such as the Patient Summary and ePrescription, but rather use the defined data structures and include collected medical information. Therefore, the NCPs developed as part of the eHDSI play an important role as a data source. How the communication between the S4H software application and NCPs will be established needs to be discussed; one possibility is to connect the NCPs to the ingestion pipeline and/or to use eDelivery.

Moreover, the terminology services could be included to ensure a proper translation of medication and other nation-specific terms.
\section{Conclusion}
\label{sec:conclusion}

This report shows that uncertainty exists regarding use cases the CEF building blocks can be applied to and open questions remain concerning legislation and usability.

For eID the identified use cases need to be validated with users; the legal question remains, whether a validation of usage criteria, such as age and citizenship, is needed. Also, not all Member States notify their electronic identification schemes, therefore not all Member States might be supported by every eID service.

Whether eSignature or eSeal is required---and which type, a simple, advanced, or a qualified one---also depends on legal questions.
Additionally, in the case of ingestion, the identification of a resource's origin, its integrity, and its encryption are conceptualized already in the SDKs; eSeal would only add value if validation of a third party is sensible.
To include the DSS for validation seems reasonable in every case.
For the DDS needs to be assessed, whether all types of signatures are supported that are intended to be used in the S4H software application.

A translation functionality is definitely needed in a cross-border scenario. For the eTranslation building block, the translation quality in the medical domain and usage conditions need to be clarified.
Clear strengths of the CEF building block are the ability to translate both plain text and formatted documents, and the vast amount of supported languages.

The eDelivery building block could be applied in multiple scenarios, however, many functionalities are already designed in form of SDKs; implementation effort and advantages of using eDelivery must be weighted carefully.

For the eHDSI, a connection to the ingestion pipeline needs to be discussed, possibly via eDelivery. Additionally, the terminology services seem promising for the translation of medical terms and medication.

This report concentrates on sample implementations that emerged under the CEF programme.
For all building blocks, commercial, eIDAS complying solutions exist.
To decide for the best alternative, these solutions need to be compared to the sample implementations regarding their suitability for S4H, usability, security, performance, and usage conditions, such as pricing.

Conclusively, it shall be noted that for a first version of the S4H software application, which will state a first iteration for user testing, the functionality of CEF facilities is not mandatory; more S4H-specific functionalities will be tested in this stage.
However, the building blocks could already be required in an early stage due to legal restrictions for electronic identification and signatures.
Open questions regarding usability should at least be discussed with citizens in first user testings.
Moreover, clear concepts of how the S4H application can be extended by CEF building blocks at a later point need to be elaborated already for an early prototype version.

\section*{Acknowledgments}

This work emerged in context of the Smart4Health project that is funded by the European Union's Horizon 2020 research and innovation programme under the grant agreement No. 826117.

\printbibliography

\end{document}